\documentclass[aps,prl,twocolumn,groupedaddress]{revtex4}                       

\usepackage{graphicx}                                                           
\usepackage{bm}                                                                 
\usepackage{amsmath}

\begin{document}

\title{Dispersion management for atomic matter waves}

\author{B. Eiermann$^1$, P. Treutlein$^{1,2}$, Th. Anker$^1$,
M. Albiez$^1$, M. Taglieber$^1$, K.-P. Marzlin$^1$, and M.K.
Oberthaler$^1$} \affiliation{ $^{1}$Fachbereich Physik, Universit{\"a}t
Konstanz,
Fach M696, 78457 Konstanz, Germany\\
$^{2}$Max-Planck-Institut f\"{u}r Quantenoptik und Sektion Physik der
Ludwig-Maximilians-Universit{\"a}t, Schellingstr.4, 80799 M\"{u}nchen,
Germany}

\date{\today}

\begin{abstract}
We demonstrate the control of the dispersion of matter wave packets
utilizing periodic potentials. This is analogous to the technique of
dispersion management known in photon optics. Matter wave packets are
realized by Bose-Einstein condensates of $^{87}$Rb in an optical dipole
potential acting as a one-dimensional waveguide. A weak optical lattice is
used to control the dispersion relation of the matter waves during the
propagation of the wave packets. The dynamics are observed in position
space and interpreted using the concept of effective mass. By switching
from positive to negative effective mass, the dynamics can be reversed.
The breakdown of the approximation of constant, as well as experimental
signatures of an infinite effective mass are studied.

\end{abstract}

\pacs{03.75.Be, 03.75.-b, 32.80.Pj, 03.75.Lm}

\maketitle

The broadening of particle wave packets due to the free space dispersion
is one of the most prominent quantum phenomena cited in almost every
textbook of quantum mechanics. The realization of Bose-Einstein
condensates of dilute gases allows for the direct observation of wave
packet dynamics in real space on a macroscopic scale \cite{BEC_general}.
Using periodic potentials it becomes feasible to experimentally study to
what extent the matter wave dispersion relation can be engineered. This
approach is similar to dispersion management for light pulses in spatially
periodic refractive index structures \cite{Agrawal01}.

First experiments in this direction have already been undertaken in the
context of Bloch oscillations of thermal atoms \cite{Dahan96} and
condensates \cite{Morsch01b}. The modification of the dipole mode
oscillation frequency of a condensate due to the changed dispersion
relation in the presence of a periodic potential has been studied in
detail \cite{Burger01,Kraemer02}. In contrast to these experiments where
the center of mass motion was studied, we are investigating the evolution
of the spatial distribution of the atomic cloud in a quasi one-dimensional
situation. For the wave packet dynamics many new effects are expected due
to the interplay between nonlinearity resulting from atom-atom interaction
and dispersion. In particular, non-spreading wave packets such as gap
solitons \cite{Meystre01} and self trapped states \cite{Trombettoni01} are
predicted. Our experiments show that even propagation in the linear regime
with modified dispersion can change the shape of an initially Gaussian
wave packet significantly.

\begin{figure}[t]
\includegraphics{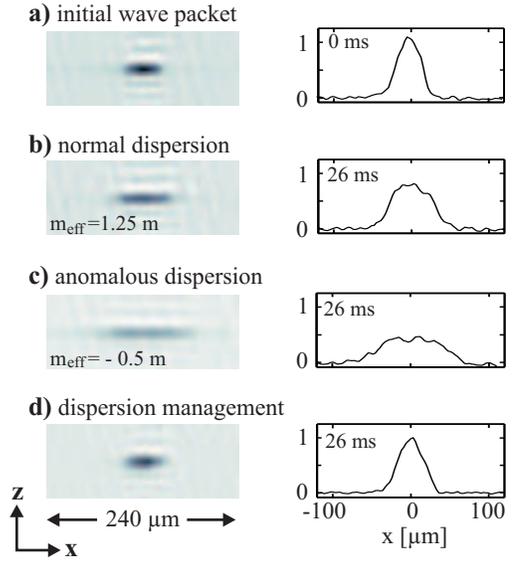}
\caption{\label{fig:focusing} Controlling the dispersion of an
atomic wave packet in a waveguide using a periodic potential.
Shown are absorption images of the wave packet  averaged over four
realizations (left) and the corresponding density distributions
$n(x,t)$ along the waveguide (right). (a) Initial wave packet.
(b,c) Images taken after an overall propagation time of
$t=26\,\text{ms}$ for different dispersion regimes with different
effective masses as indicated. (d) Wave packet subjected to
dispersion management: an initial stage of expansion for
$t=17\,\text{ms}$ with normal dispersion is followed by
propagation with anomalous dispersion for $t=9\,\text{ms}$. The
broadening in the normal dispersion regime has been reversed by
anomalous dispersion.}
\end{figure}

For atomic matter waves inside a one-dimensional optical waveguide, we
have achieved dispersion management by applying a weak periodic potential
with adjustable velocity. Fig.~\ref{fig:focusing} shows the results of an
experiment in which the propagation of an atomic wave packet is studied in
the normal (Fig.~\ref{fig:focusing}b) and anomalous
(Fig.~\ref{fig:focusing}c) dispersion regime corresponding to positive and
negative effective mass, respectively.
A broadening of the wave packet is observed in both cases. The
faster spreading in the case of anomalous dispersion is a
consequence of the smaller absolute value of the negative
effective mass. However, if we switch from one regime to the other
during the propagation by changing the velocity of the periodic
potential, the effects of normal and anomalous dispersion cancel.
The wave packet, which has initially broadened under the influence
of normal dispersion, reverses its expansion and compresses until
it regains its initial shape (Fig.~\ref{fig:focusing}d). This is a
direct proof of the realization of negative effective mass.

\begin{figure}[t]
\includegraphics{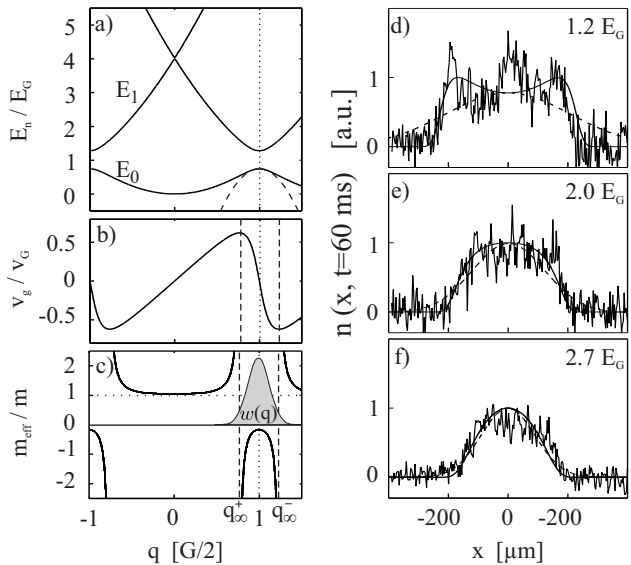}
\caption{\label{fig:shapes} (a) Band structure in the first Brillouin zone
for atoms in an optical lattice with $V_0=1.2\,E_G$ (solid), parabolic
approximation to the lowest energy band at $q=G/2$ (dashed), corresponding
group velocity (b) and effective mass (c) in the lowest energy band. The
vertical dashed lines at $q=q^\pm_\infty$ indicate where
$|m_\mathrm{eff}|=\infty$. (d-f) Spatial densities of the wave packet
after $t=60\,\text{ms}$ of propagation with $q_c=G/2$ for different $V_0$.
The position $x$ along the waveguide is measured in the moving frame of
the optical lattice. The solid lines represent the theoretical predictions
using linear propagation with the exact band structure and the
quasimomentum distribution given in graph (c). The dashed lines in Graph
(d-f) represent the prediction of the constant effective mass
approximation.}
\end{figure}

The concept of effective mass $m_\mathrm{eff}$
\cite{Ashcroftenglish76} allows to describe the
 dynamics of matter wave packets inside a periodic
potential in a simple way via a modified dispersion relation. The
periodic potential in our experiments is well described by
\begin{equation*}
V(x)=\frac{V_0}{2} \cos \left( G x \right)
\end{equation*}
with a modulation depth $V_0$ on the order of the grating recoil energy
$E_G=\hbar^2 G^2/8m$, with $G=2 \pi/d$ where $d=417\,\text{nm}$ represents
the spatial period. The energy spectrum of atoms inside the periodic
potential exhibits a band structure $E_n(q)$ which is a periodic function
of quasimomentum $q$ with periodicity $G$ corresponding to the width of
the first Brillouin zone (Fig.~\ref{fig:shapes}a). In our experiment, we
prepare condensates  in the lowest energy band ($n=0$) with a
quasimomentum distribution $w(q)$ centered at $q=q_c$ with an rms width
$\Delta q \ll G$ \cite{Denschlag}.

It has been shown by Steel et al. \cite{Steel98} that in this case the
condensate wave function in a quasi one-dimensional situation can be
described by $\Psi(x,t)=A(x,t)\,\,\phi_{q_c}(x)\,\,exp(-i
E_0(q_c)t/\hbar)$, where $\phi_{q_c}$ represents the Bloch function in the
lowest energy band corresponding to the central quasimomentum. The
evolution of the envelope function $A(x,t)$, normalized to the total
number of atoms $N_0$, is described by
\begin{equation}\label{gl:SGL}
i \hbar \left(\frac{\partial}{\partial t}
+v_g\frac{\partial}{\partial x}\right) A = - \frac{\hbar^2}{2
m_\mathrm{eff}} \frac{\partial^2}{\partial x^2} A+\tilde{g}|A|^2
A.
\end{equation}
The strength of the atom-atom interaction is given by $\tilde{g}=
\alpha_{nl} 2 \hbar \omega_\perp a$, with the transverse trapping
frequency of the waveguide $\omega_\perp$, the s-wave scattering length
$a$, and a renormalization factor $\alpha_{nl}= 1/d \int_{0}^{d} dx
|\phi_{q_c}|^4$. Besides the modification of the nonlinear term the
periodic potential leads to a group velocity of the envelope $A(x,t)$
determined by the energy band via $v_g(q_c)=\hbar^{-1}(\partial
E_0(q)/\partial q)|_{q_c}$ (Fig.~\ref{fig:shapes}b). In addition, the
kinetic energy term describing the dispersion of the wave packet is
modified by the effective mass (Fig.~\ref{fig:shapes}c)
\begin{equation*}
m_\mathrm{eff}(q_c)=\hbar^2 \left( \left.\frac{\partial^2 E_0(q)
}{\partial q^2} \right|_{q_c} \right)^{-1}.
\end{equation*}
It is important to note that group velocity and effective mass are assumed
to be constant during the propagation. Since the approximation of constant
effective mass corresponds to a parabolic approximation of the energy
band, it is only valid for sufficiently small $\Delta q$ as can be seen in
Fig.~\ref{fig:shapes}.

The general solution of Eq.$\,$(\ref{gl:SGL}) is a difficult task, but
simple analytic expressions can be found in the special cases of
negligible and dominating atom-atom interaction. Omitting the last term in
Eq.~(\ref{gl:SGL}) it is straightforward to see that $|m_\mathrm{eff}|$
controls the magnitude of the dispersion term and thus the timescale of
the wave packet broadening. A change in sign of $m_\mathrm{eff}$
corresponds to time reversal of the dynamics in a frame moving with
velocity $v_g$. In the regime where the atom-atom interaction is
dominating, e.g. during the initial expansion of a condensate, the
evolution of the envelope function can be found in standard nonlinear
optics textbooks \cite{Agrawal95} and in form of scaling solutions in the
context of Bose-Einstein condensates \cite{scaling}. Note that in this
regime the kinetic energy term is still relevant and thus a change of the
sign of the effective mass will reverse the dynamics.

In the following we will analyze the obtained experimental results in more
detail. The initial wave packet shown in Fig.~\ref{fig:focusing}a) is
characterized by $\Delta x_0=14.8(6) \,\mu\text{m}$ ($\Delta x$ is the
r.m.s width of a Gaussian fit) and shows the density distribution of a
Bose-Einstein condensate of $2\cdot10^4$ atoms in a three-dimensional
dipole trap. Before releasing the atomic cloud into the one dimensional
waveguide, a weak periodic potential along the waveguide is adiabatically
ramped up to $V_0=2.8(2) \,E_G$ within $6\,\text{ms}$. This turns the
initial Gaussian momentum distribution of the atoms into a Gaussian
distribution of quasimomenta $w(q)$ centered at $q_c=0$ in the lowest
energy band of the lattice with a corresponding effective mass
$m_\mathrm{eff}=1.25\,m$. The density distribution shown in
Fig.~\ref{fig:focusing}b) is a result of propagation within the stationary
periodic potential for $t=26\, \text{ms}$ and exhibits a spread of $\delta
x:=\sqrt{\Delta x(t)^2-\Delta x_0^2} = 18.4(12) \mu$m in contrast to
$\delta x_f=20.2(14)\,\mu\text{m}$ for expansion without periodic
potential. The resulting ratio $\delta x_f/\delta x =1.10(15)$ indicates
that the evolution is dominated by the nonlinearity, in which case one
expects $\delta x_f/\delta x \approx \sqrt{m_\mathrm{eff}/m}=1.11$
\cite{Potasek86}. In the case of linear propagation one expects $\delta
x_f/\delta x = m_\mathrm{eff}/m=1.25$.

The dynamics in the anomalous dispersion regime are investigated by
initially accelerating the periodic potential within $3\,\text{ms}$ to a
velocity $v=v_G:=\hbar G/2 m$, thus preparing the atomic wave packet at
the edge of the Brillouin zone, where $m_\mathrm{eff}=-0.5\,m$. The
velocity is kept constant during the subsequent expansion. In the regime
of negative  mass a condensate exhibits collapse dynamics similar to
condensates with attractive atom-atom interaction. Two-dimensional
calculations for our experimental situation reveal that this collapse
happens within the initial 3-6$\,$ms of propagation. This leads to a fast
reduction of the density and therefore of the nonlinearity due to
excitation of transverse states. An indication of the population of
transverse states is the observed increase of the transverse spatial
extension of the wave packets for anomalous dispersion by almost a factor
of two. The optical resolution of our setup does not allow for a
quantitative analysis of the transverse broadening. The measured spread of
$\delta x =38.5(15)\,\mu \text{m}$ after $23\,\text{ms}$ of expansion
leads to a ratio $\delta x_f/\delta x= 0.46(5)$. This ratio represents an
upper limit due to the initial collapse, indicating that a coherent
nonlinear evolution in one dimension cannot describe the experimental
results.

In the case of dispersion management Fig.~\ref{fig:focusing}d) the wave
packet was first subjected to normal dispersion for $17\,\text{ms}$ at
$q_c=0$. The time of subsequent propagation with anomalous dispersion at
$q_c=G/2$ was adjusted to achieve the minimal wave packet size of $\Delta
x= 15.4(2)\,\mu \text{m}$. The minimum was achieved for times ranging from
$7\, \text{ms}$ to $9\, \text{ms}$ which is in rough agreement with the
expected time resulting from effective mass considerations
$\sqrt{0.5/1.25} \times 17\,\text{ms} = 10.7\,\text{ms}$. The experimental
results exhibit a rapid increase of the wave packet width for overall
propagation times longer than $27\,\text{ms}$ also indicating collapse
dynamics.

Before presenting the detailed experimental studies of the
applicability of the constant effective mass approximation and
evidence of infinite masses, we turn to a brief description of our
setup.

The wave packets have been realized with a $^{87}$Rb Bose-Einstein
condensate. The atoms are collected in a magneto-optical trap and
subsequently loaded into a magnetic time-orbiting potential trap. By
evaporative cooling we produce a cold atomic cloud which is then
transferred into an optical dipole trap realized by two focused Nd:YAG
laser beams with $60\,\mu\text{m}$ waist crossing at the center of the
magnetic trap. Further evaporative cooling is achieved by lowering the
optical potential leading to pure Bose-Einstein condensates with up to $3
\cdot 10^4$ atoms in the $|F=2, m_F=+2\rangle$ state. By switching off one
dipole trap beam the atomic matter wave is released into a trap acting as
a one-dimensional waveguide with radial trapping frequency $\omega_\perp =
2 \pi \cdot 80 \,\text{Hz}$ and longitudinal trapping frequency
$\omega_\parallel = 2 \pi \cdot 1.5 \,\text{Hz}$.

The periodic potential is realized by a far off-resonant standing
light wave with a single beam peak intensity of up to $5\,\text
{W/cm}^2$. The chosen detuning of 2\,nm to the blue off the D2
line leads to a spontaneous emission rate below $1\, \text{Hz}$.
The frequency and phase of the individual laser beams are
controlled by acousto-optic modulators driven by a two channel
arbitrary waveform generator allowing for full control of the
velocity and amplitude of the periodic potential. The light
intensity and thus the absolute value of the potential depth was
calibrated independently by analyzing results on Bragg scattering
\cite{Bragg} and Landau Zener tunneling \cite{Morsch01b}.

The wave packet evolution inside the combined potential of the
waveguide and the lattice is studied by taking absorption images
of the atomic density distribution after a variable time delay.
The density profiles along the waveguide, $n(x,t)$, are obtained
by integrating the absorption images over the transverse dimension
$z$.

Since the assumption of a constant effective mass used so far is only an
approximation, it is important to check its applicability in experiments.
Therefore we investigate the dynamics of wave packets prepared at the
Brillouin zone edge ($q_c=G/2$) for different potential depths. The
observed density profiles after $60\,\text{ms}$ of propagation are shown
in Fig.~\ref{fig:shapes}d-f). While both the initial wave packet shape
$n(x,0)$ and the quasimomentum distribution $w(q)$ are measured to be
approximately Gaussian, the wave packet changes its shape during
evolution. We attribute this distortion to the invalidity of the constant
effective mass approximation, which assumes that the populated
quasimomenta experience the same negative curvature of $E_0(q)$. Since the
range of quasimomenta fulfilling this criterion becomes smaller with
decreasing modulation depth, a more pronounced distortion of the wave
packet shape  for weak potentials is expected (see
Fig.~\ref{fig:shapes}d).

This explanation is confirmed more quantitatively by comparing the
observed wave packets with the results of an integration of
Eq.$\,$(\ref{gl:SGL}) neglecting the nonlinear term. Since the initial
collapse of the condensate cannot be described by a linear theory, we take
a Gaussian function fitted to the density distribution measured at
$20\,\text{ms}$ as the initial wave packet for the numerical propagation.
Due to the fact that this is not a minimum uncertainty wave packet we add
a quadratic phase in real space such that the Fourier transform of the
wave packet is consistent with the measured momentum distribution. In
first approximation this takes into account the initial expansion
including the repulsive atom-atom interaction. For the subsequent
propagation of $40\, \text{ms}$ we use the full expression for $E_0(q)$
which is obtained numerically. Finally we convolute the obtained density
distribution with the optical resolution of our setup. In
Fig.~\ref{fig:shapes}d-f) we compare the data with the linear theory
described above (solid line) and with the constant effective mass
approximation (dashed line). Clearly the effective mass approximation
cannot explain the observed distortion and it strongly overestimates the
expansion velocity for weak potentials. Additionally, for small potential
modulation depths new features appear in the central part of the wave
packet which cannot be explained using the linear theory. We are currently
investigating these features in more detail.

The observed distortion is mainly a consequence of another very
interesting feature of the band structure: the existence of
$|m_\mathrm{eff}|=\infty$ for certain quasimomenta $q=q^\pm_\infty$ (see
Fig.~\ref{fig:shapes}c). A diverging mass implies that the group velocity
is extremal and the dispersion vanishes as can be seen from
Eq.$\,$(\ref{gl:SGL}). As a consequence an atomic ensemble whose
quasimomentum distribution is overlapping $q=q^\pm_\infty$ will develop
steep edges as can be seen in Fig.~\ref{fig:shapes}d). These edges
propagate with the maximum group velocity of the lowest band.

\begin{figure}
\includegraphics{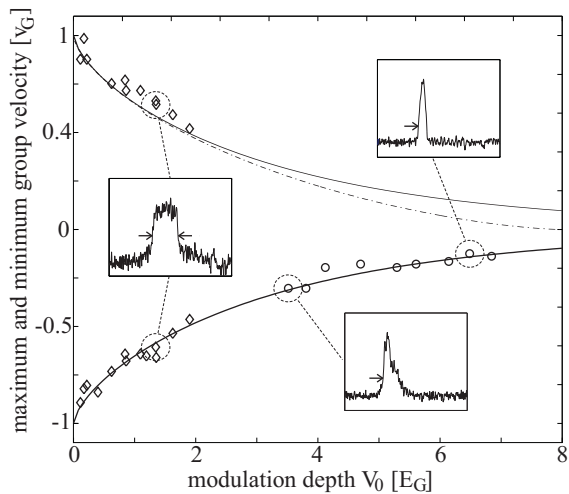}
\caption{\label{fig:vmax} Group velocities of steep edges emerging
from an initial wave packet with significant overlap with
$q^\pm_\infty$ in quasimomentum space. The measured velocities of
the indicated positions (arrows in insets) agree very well with
the expected maximum and minimum velocity in the lowest band
(solid line) corresponding to the infinite masses. The dashed line
represents the prediction of the weak potential approximation
\cite{Ashcroftenglish76}. For potentials smaller than 2 $E_G$
(diamonds) data are obtained by preparing the initial wave packet
at $q=G/2$ leading to two steep edges (see inset). For higher
potentials (circles) the wave packet is prepared at $q = G/4$ to
ensure population of the quasimomentum corresponding to infinite
mass.}
\end{figure}
The systematic investigation of the velocities of the edges is shown in
Fig.~\ref{fig:vmax} for different values of $V_0$. In order to get a
significant overlap of $w(q)$ with $q=q^\pm_\infty$, we prepare atomic
ensembles with $\Delta q=0.17\,G/2 $ at $q_c=G/2$ realized by
Bose-Einstein condensates of $2 \cdot 10^4$ atoms with a spatial extension
of $\Delta x = 15\,\mu\text{m}$. The velocities of the edges are derived
from two images taken at $20\,\text{ms}$ and $60\,\text{ms}$,
respectively. In each image the position of the edge is evaluated at the
levels indicated by the arrows in the insets of Fig.~\ref{fig:vmax} (50\%
and 25\% of the maximum density). Since the momentum spread is too small
to populate the infinite mass points for potentials deeper than $2 E_G$
the atomic ensemble was then prepared at $q_c=G/4$ by accelerating the
periodic potential to the corresponding velocity $v=v_G/2$. The resulting
wave packet shapes are asymmetric exhibiting a steep edge on one side
which becomes less pronounced for potentials deeper than $5\,E_G$. The
obtained experimental results in Fig.~\ref{fig:vmax} are in excellent
agreement with the numerically calculated band structure predictions. In
contrast to the good agreement of the maximum velocity for all potential
depths we find that for $V_0>5\,E_G$ the group velocity of the center of
mass is only 10\% of the expected velocity deduced from the dispersion
relation at $q_c=G/4$. This could be an indication of entering the tight
binding regime where the nonlinear effect of self trapping, i.e. stopping
and non-spreading wave packets, has been predicted by Trombettoni et al.
\cite{Trombettoni01}. We are currently investigating the transport
properties in this regime in more detail.

In conclusion, we have demonstrated experimentally that the
dispersion of atomic matter waves in a wave\-guide can be
controlled using a weak periodic potential. Matter wave packets
with positive, negative and infinite effective masses are studied
in the regime of weak and intermediate potential heights. The
preparation of matter waves with engineered dispersion
($m_\mathrm{eff}<0$) is an important prerequisite for the
experimental investigation of atomic gap solitons and other
effects arising from the coherent interplay of nonlinearity and
dispersion in periodic potentials.

We wish to thank J. Mlynek for his generous support, A. Sizmann and B.
Brezger for many stimulating discussions, and J. Bellanca, and K.
Forberich for their help in building up the experiment. This work was
supported by Deutsche Forschungsgemeinschaft, Emmy Noether Program, and by
the European Union, Contract No. HPRN-CT-2000-00125.


\begin{thebibliography}{10}

\bibitem{BEC_general}
{\itshape Bose-Einstein condensation in atomic gases}, ed. by
M.~Inguscio, S.~Stringari,and C. Wieman, (IOS Press, Amsterdam
1999)

\bibitem{Agrawal01}
G.P.~Agrawal, {\itshape Applications of Nonlinear Fiber Optics}
(Academic Press, San Diego, 2001).


\bibitem{Dahan96}
M.~Ben~Dahan, E.~Peik, J.~Reichel, Y.~Castin, and C.~Salomon,
Phys.Rev.Lett. {\bf 76} 4508 (1996).

\bibitem{Morsch01b}
O.~Morsch, J.~M\"uller, M.~Cristiani, D.~Ciampini, and
E.~Arimondo, Phys. Rev. Lett. {\bfseries 87}, 140402 (2001).

\bibitem{Burger01}
S.~Burger, F.S.~Cataliotti, C.~ Fort, F.~Minardi, M.~Inguscio,
M.L.~Chiofalo, and M.P.~Tosi, Phys. Rev. Lett. {\bfseries 86},
4447 (2001).

\bibitem{Kraemer02}
M.~Kr\"amer, L.~Pitaevskii, and S.~Stringari, Phys. Rev. Lett
{\bfseries 88}, 180404 (2002).


\bibitem{Meystre01} P.~Meystre, {\itshape Atom Optics} (Springer Verlag, New York,
2001) p 205, and references therein.

\bibitem{Trombettoni01}
A.~Trombettoni and A.~Smerzi, Phys. Rev. Lett. {\bfseries 86},
2353 (2001).

\bibitem{Ashcroftenglish76}
N.~Ashcroft and N.~Mermin, {\itshape Solid State Physics}
(Saunders, Philadelphia, 1976).

\bibitem{Denschlag}
J.~Hecker-Denschlag, J.E.~Simsarian , H.~H\"affner, C.~McKenzie, A.~
Browaeys, D.~Cho, K.~Helmerson, S.L.~Rolston and W.D.~Phillips, J. Phys. B
{\bfseries 35}, 3095 (2002), and references therein.


\bibitem{Steel98}
M.~Steel and W.~Zhang, cond-mat/9810284 (1998).


\bibitem{Agrawal95}
G.~Agrawal, {\itshape Nonlinear Fiber Optics} (Academic Press, San
Diego, 1995).


\bibitem{scaling}
Y.~Castin, and R.~Dum, Phys.Rev.Lett. {\bfseries 77}, 5315 (1996);
Y.~Kagan, E.L. Surkov, and G.V.~Shlyapnikov, Phys.Rev. A {\bfseries 54}
R1753 (1996)

\bibitem{Potasek86}
M.J.~Potasek, G.P.~Agrawal, S.C.~Pinault, J.Opt.Soc.Am.~B
{\bfseries 3}, 205 (1986).

\bibitem{Bragg}
M.~Kozuma, L.~Deng, E.W.~Hagley, J.~Wen, R.~Lutwak, K.~Helmerson,
S.L.~Rolston, and W.D.~Phillips, Phys.Rev.Lett. {\bfseries 82} 871
(1999).

\end{thebibliography}
\end{document}